\documentclass[final,5p,times,twocolumn]{elsarticle}

\usepackage{hyperref}
\hypersetup{colorlinks=true, citecolor=blue, filecolor=blue, linkcolor=blue, urlcolor=blue}
\usepackage[utf8]{inputenc}
 
\usepackage{listings}
\usepackage{xcolor}
 
\definecolor{codegreen}{rgb}{0,0.6,0}
\definecolor{codegray}{rgb}{0.5,0.5,0.5}
\definecolor{codepurple}{rgb}{0.58,0,0.82}
\definecolor{backcolour}{rgb}{0.95,0.95,0.92}
 
\lstdefinestyle{mystyle}{
    backgroundcolor=\color{backcolour},   
    commentstyle=\color{codegreen},
    keywordstyle=\color{magenta},
    numberstyle=\tiny\color{codegray},
    stringstyle=\color{codepurple},
    basicstyle=\ttfamily\footnotesize,
    breakatwhitespace=false,         
    breaklines=true,                 
    captionpos=b,                    
    keepspaces=true,                 
    numbersep=5pt,                  
    showspaces=false,                
    showstringspaces=false,
    showtabs=false,                  
    tabsize=2
}
 
\usepackage{amssymb}

\newcounter{bla}

\journal{Computer Physics Communications}

\begin{document}

\begin{frontmatter}



\title{PyXtal: a Python Library for Crystal Structure Generation and Symmetry Analysis}


\author[a]{Scott Fredericks}
\author[a]{Kevin Parrish}
\author[a]{Dean Sayre}
\author[a]{Qiang Zhu\corref{author}}

\cortext[author] {Corresponding author.\\\textit{E-mail address:} qiang.zhu@unlv.edu}
\address[a]{Department of Physics and Astronomy, University of Nevada Las Vegas, Las Vegas, Nevada 89154, USA}

\begin{abstract}
We present PyXtal, a new package based on the Python programming language, used to generate structures with specific symmetry and chemical compositions for both atomic and molecular systems. This software provides support for various systems described by point, rod, layer, and space group symmetries. With only the inputs of chemical composition and symmetry group information, PyXtal can automatically find a suitable combination of Wyckoff positions with a step-wise merging scheme. Further, when the molecular geometry is given, PyXtal can generate different dimensional organic crystals with molecules occupying both general and special Wyckoff positions. Optionally, PyXtal also accepts user-defined parameters (e.g., cell parameters, minimum distances and Wyckoff positions). In general, PyXtal serves three purposes: (1) to generate custom structures, (2) to modulate the structure by symmetry relations, (3) to interface the existing structure prediction codes that require the generation of random symmetric structures. In addition, we provide several utilities that facilitate the analysis of structures, including symmetry analysis, geometry optimization, and simulations of powder X-ray diffraction (XRD). Full documentation of PyXtal is available at \url{https://pyxtal.readthedocs.io}.  

\end{abstract}

\begin{keyword}
Symmetry; Crystallography; Structure prediction; Wyckoff sites; Global optimization; Phase transition
\end{keyword}

\end{frontmatter}

\noindent
{\bf PROGRAM SUMMARY}\\
\begin{small}
\noindent
{\em Program Title:} PyXtal \\
{\em Licensing provisions:} MIT \cite{1}\\
{\em Programming language:} Python 3 \\
{\em Nature of problem:} Knowledge of structure at the atomic level is the key to understanding materials' properties. Typically, the structure of a material can be determined either from experiment (such as X-ray diffraction, spectroscopy, microscopy) or from theory (e.g., enhanced sampling, structure prediction). In many cases, the structure needs to be solved iteratively by generating a number of trial structure models satisfying some constraints (e.g., chemical composition, symmetry, and unit cell parameters). Therefore, it is desirable to have a computational code that is able to generate such trial structures in an automated manner.\\
{\em Solution method:} The PyXtal package is able to generate many possible random structures for both atomic and molecular systems with all possible symmetries. To generate the trial structure, the algorithm can either start with picking the symmetry sites randomly from high to low multiplicities, or use sites that are predefined by the user. For molecules, the algorithm can automatically detect the molecules' symmetry and place them into special Wyckoff positions while satisfying their compatible site symmetry. With the support of symmetry operations for point, rod, layer and space groups, PyXtal is suitable for the computational modeling of systems from zero, one, two, and three dimensional bulk crystals. \\
\  
   \\

\end{small}

\section{Introduction}
\label{intro}

Knowing the atomic structure is the key to understanding the properties of materials. Ideally, the full atomic structure can be experimentally determined through single crystal X-ray diffraction. If a single crystal sample is not available, only partial structural information can be extracted from various characterizations, such as powder X-ray diffraction/absorption, Raman spectroscopy, nuclear magnetic resonance, and electron microscopy. Based on this partial structural information (e.g., symmetry, unit cell), a number of trial structures are constructed and optimized at their corresponding thermodynamic conditions. The simulated pattern for each relaxed structure is then compared with the observed one. By doing this iteratively, the structure can be finally resolved. It has been previously demonstrated that structures can be predicted computationally using first-principles \cite{Oganov-Book-2011, Oganov2019}. The basic idea of computational structure prediction is to guess the correct crystal structure under specific conditions by computationally sampling a wide range of possible structures via different global optimization techniques (e.g., random search \cite{Pickard-JPCM-2011}, metadynamics \cite{Martonak-PRL-2003}, basin hopping \cite{LJClusters}, evolutionary algorithms \cite{Oganov-JCP-2006, XtalOpt}, particle swarm optimization \cite{Wang-PRB-2010}). After many attempts, the most energetically stable structure found is the one most likely to exist.

For structure determination from either partial experimental information or pure computation, a number of trial structures are needed. It is generally believed that by beginning with already-symmetric structures, fewer attempts are needed to find the global energy minimum \cite{Pickard-JPCM-2011, Lyakhov-CPC-2013}. For inorganic crystals, symmetry constraints have been encoded in many computational structure prediction codes such as AIRSS \cite{Pickard-JPCM-2011}, USPEX \cite{Lyakhov-CPC-2013}, CALYPSO \cite{CALYPSO} and XtalOpt \cite{XtalOpt}. For a given crystal with symmetry, the atomic positions are classified by Wyckoff positions (WP) \cite{1.2}. Two approaches are used to place the atoms into the Wyckoff sites so that the structure satisfies the desired symmetry. One is to pre-generate a set of WPs and then add atoms to these sites \cite{CALYPSO, randspg, Burkhard-PRB-2018}. The other is to place atoms to the most general WPs and then merge them to the special sites if there exist close atomic pairs \cite{Lyakhov-CPC-2013, zhu2012}. This will be repeated until the desired stoichiometry is achieved. The development of new computational tools has allowed the structures of many new and increasingly complex materials to be anticipated \cite{Oganov2019}.

For the prediction of organic crystals, the role of symmetry is even more pronounced. In the periodically conducted Blind Tests of organic crystal structure prediction organized by the Cambridge Crystallographic Data Centre \cite{reilly:2016:6th_blind_test}, most research groups attempted to reduce the structure generation to a limited range of space group choices with one molecule in the asymmetric unit ($Z^\prime$). This is based on a statistical analysis that most organic crystals tend to crystallize in only a few space groups with $Z^\prime$ = 1 \cite{Baur-1992}. Currently, there exist a few free packages \cite{upack, molpack} which allow the generation of random molecular crystals with $Z^\prime$ = 1. Combined with modern structure search algorithms (including quasi-random \cite{Case-JCTC-2016}, parallel tempering \cite{Neumann-ANIE-2008}, genetic algorithms \cite{Curtis-JCTC-2018}, and evolutionary methods \cite{zhu2012}), one can perform an extensive search for the plausible structures. Their energies can then be evaluated with different energy models from the empirical to ab-initio level. A recent blind test \cite{reilly:2016:6th_blind_test} has shown that the combination of effective structure generation and an energy ranking scheme can predict not only the structure of simple rigid molecules, but also the molecules representing real-life challenges. 

Despite the fact that many programs have their own built-in functions to generate crystals with specific space groups or clusters with specific point groups, most of these functions are implemented in the main packages and cannot work in a standalone manner. To our knowledge, there is only one open source code Randspg \cite{randspg} that provides the interface to generate 3D atomic crystal structures. Similarly, most molecular crystal generators only support molecules occupying the general WPs with $Z^\prime$ = 1, except for the recent development of Genarris 2.0 \cite{tom2020genarris} which is able to deal with structures having a non-integer value of $Z^\prime$ (meaning molecules can occupy special WPs). So far, there is no single code which enables the generation of molecular crystals with arbitrary $Z^\prime$, varying from fractional to multiple integers. While 90\% of organic crystals in the Cambridge Structure Database (CSD) have $Z^\prime$=1, recent advances in experimental polymorph searching and crystal engineering highlight the rich variety of multi-component crystals (co-crystals, salts, solvates, etc) as well as crystal structures with multiple molecules in the asymmetric unit. For instance, many well-studied molecules, including aspirin \cite{Shtukenberg-CGD-2017}, resorcinol \cite{zhu:2016:resorscinol}, coumarin \cite{Shtukenberg-CC-2017}, glycine \cite{Xu-ANIE-2017}, DDT \cite{kahr:2017:DDT_polymorphs}, and ROY \cite{Tan-FD-2018}, were found to adopt crystal structures with $Z^\prime>1$. Lastly, neither Randspg nor Genarris supports the generation of low dimensional crystals, which require explicit consideration of layer/rod/point-group (instead of space-group) symmetry operations. Collectively, these cases motivated us to develop a standalone Python program called PyXtal which can be used for customized structure generation for different-dimensional systems, including atomic clusters and 1D/2D/3D atomic/molecular crystals. In Sections \ref{algo} and \ref{dependence}, we will detail the algorithms and the software dependencies. The basic usages of PyXtal will be introduced in Section \ref{usage}, followed by two example studies using PyXtal in the context of structure prediction in Section \ref{examples}. Finally, we summarize the features of PyXtal and conclude the manuscript in Section \ref{conclusion}.


\section{Algorithms}
\label{algo}
The core algorithms in PyXtal involve (1) generation of random symmetric crystals and (2) modulation of structures according to the symmetry relation.

\subsection{Structure Generation}
\begin{figure}[!htbp]
    \centering
    \includegraphics[width=0.25\textwidth]{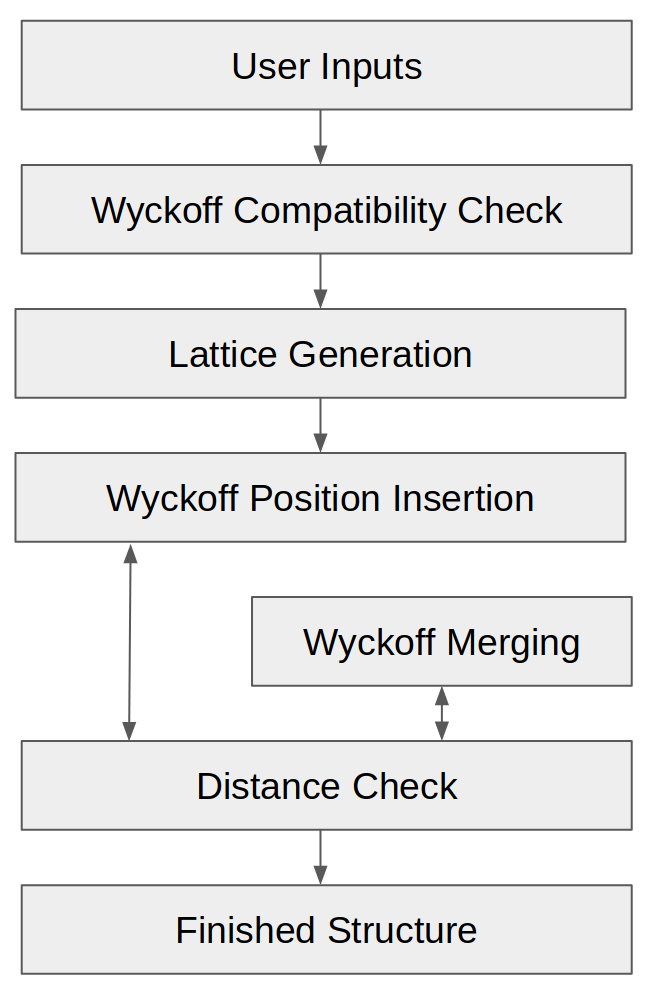}
    \caption[PyXtal structure generation flowchart]{PyXtal structure generation flowchart. Generation is based on inputs from the user.}
    \label{fig:Flowchart}
\end{figure}

PyXtal adopts the following algorithm to generate a trial structure. First, the user inputs their choice of dimension (0, 1, 2, or 3), symmetry group, stoichiometry, and relative volume of the unit cell. Optionally, additional parameters may be chosen that constrain the unit cell and maximum inter-atomic distance tolerances. Next, PyXtal checks if the stoichiometry is compatible with the choice of symmetry group. If the check passes, trial structure generation begins. Figure \ref{fig:Flowchart} shows a flowchart of the algorithm.

Each step has a maximum number of attempts. If the generation attempt fails at any point, the algorithm will revert progress for the current step and try again until the maximum limit of attempts is encountered. This ensures that the algorithm stops in a reasonable amount of time while still giving each generated parameter a chance for success. For certain inputs, structure generation may take many attempts or fail after the maximum number of attempts. Typically, these failures indicate that the input parameters are not likely to produce a realistic structure without fine-tuning the atomic positions. In such cases, a larger unit cell volume or a smaller distance tolerance may prevent failure. Below we discuss the technical details implemented during structure generation.

\subsubsection{Wyckoff Compatibility Checking}
Before generating a trial structure, PyXtal performs a WP compatibility check. Since WPs in different space groups have different multiplicities, this is a required step that ensures compatibility between a stoichiometry and its assigned space group. For example, consider the space group \textit{Pn-3n} (\#222), which has a minimum WP of 2a, followed by 6b. To create a crystal structure with 4 atoms in the unit cell for this symmetry group, the combination of Wyckoff positions must add up to 4. Here, this is not possible. The position 2a cannot be repeated, because it falls on the exact coordinates (1/4, 1/4, 1/4) and (3/4, 3/4, 3/4). A second set of atoms in the 2a position would overlap the atoms in the first position, which is physically impossible.

Thus, from our previous discussion, it is necessary to check the input stoichiometry against the WPs of the desired space group. PyXtal implements this by iterating through all possible combinations of WPs within the confinements of the given stoichiometry. As soon as a valid combination is found, the check returns True. Otherwise, if no valid combination is found, the check returns False and the generation attempt raises a warning.

Some space groups allow valid combinations of WPs, but do not permit many (or any) positional degrees of freedom within the structure. It may also be the case that the allowed combinations result in atoms that are too close together. In these cases, PyXtal will attempt generation as usual: it will continue to search for a compatible structure until the maximum limit is reached, or until a successful generation occurs. In the event that structure generation repeatedly fails for a given combination of space group and stoichiometry, the user should make note and avoid the combination going forward.

\subsubsection{Lattice Generation}

The first step in PyXtal's structure generation is the choice of unit cell. Depending on the symmetry group, a specific type of lattice must be generated. For all crystals, the conventional cell choice is used to avoid ambiguity. Lattice information can be pre-defined by the user in either vector form ($a$, $b$, $c$, $\alpha$, $\beta$, $\gamma$) or in the form of a 3$\times$3 matrix. If lattice information is not provided, PyXtal will attempt to estimate the volume based on the chemical composition, resulting in the generation of a random unit cell which satisfies the input constraints.

The most general case is the triclinic cell, from which other cell types can be obtained by applying certain constraints. To generate a triclinic cell, 3 real numbers are randomly chosen (using a Gaussian distribution centered at 0) as the off-diagonal values for a 3$\times$3 shear matrix. By treating this shear matrix as a cell matrix, one obtains 3 lattice angles. For the lattice vector lengths, a random 3-vector between (0, 0, 0) and (1, 1, 1) is chosen (using a Gaussian distribution centered at (0.5, 0.5, 0.5)). The relative values of the x, y, and z coordinates are used for a, b, and c respectively and scaled based on the required volume. For other cell types, any free parameters are obtained using the same methods as for the triclinic case, and then constraints are applied. In the tetragonal case, for example, all angles will be fixed to 90 degrees. Thus, only a random vector is needed to generate the lattice constants.

For low-dimensional systems, not all three unit cell axes are periodic. Therefore, the algorithm must be altered slightly, as described below.

For the 2D case, we chose $c$ to be the non-periodic axis by default. For layer groups 3-7 ($P112$, $P11m$, $P11a$, $P112/m$, $P112/a$), $c$ is also the unique axis; for all other layer groups, $a$ is the unique axis. The length of $c$ (the crystal's ``thickness") is an optional parameter which can be specified by the user. If no thickness is given, the algorithm will automatically compute a random value based on a Gaussian distribution centered at the cubic root of the estimated volume. In other words, $c$ will have the same length as the other axes on average.

For the 1D case, $c$ is the periodic axis by default. For rod groups 3-7 ($P221$, $Pm11$, $Pc11$, $P2/m11$, $P2/c11$), $a$ is the unique axis; for all other rod groups, $c$ is the unique axis. Instead of choosing a value for the thickness, we constrain the unit cell based on the cross-sectional area of the \textit{a-b} plane. This area can be either specified by the user or generated randomly. As with the 2D and 3D cases, there is no preference for any axis to be longer or shorter than the others unless specified by the user.

For 0D clusters, we constrain the atoms to lie within either a sphere or an ellipsoid, depending on the point group. For spherically or polyhedrally symmetric point groups ($C_1$, $C_i$, $D_2$, $D_{2h}$, $T$, $T_h$, $O$, $T_d$, $O_h$, $I$, $I_h$), we define a sphere centered on the origin. For all other point groups (which have a unique rotational axis), we define an ellipsoid with its $c$-axis aligned with the rotational axis. The $a$- and $b$-axes are always of equal length to ensure rotational symmetry about the $c$-axis. The relative lengths for the ellipsoidal axes are chosen in the same way as for the 3D tetragonal case. In order for the 0D case to be compatible with the 1D, 2D, and 3D cases, we encode the spheres and ellipsoids as lattices (a cubic lattice for a sphere, or tetragonal lattice for an ellipsoid). Then, when generating atomic coordinates, we check whether the randomly chosen point lies within the sphere or ellipsoid. If not, we simply retry until it does.

\subsubsection{Wyckoff Position Selection and Merging}

The central building block for crystals in PyXtal is the WP. Once a space group and lattice are chosen, WPs are inserted one at a time to add structure. In PyXtal, we closely follow the algorithm provided in Ref. \cite{Lyakhov-CPC-2013} to place the atoms in different WPs. In general, PyXtal starts with the largest available WP, which is the general position of the symmetry group. If the number of atoms required is equal to or greater than the size of the general position, the algorithm proceeds. If fewer atoms are needed, the next largest WP (or set of WPs) is chosen, in order of descending multiplicity. This is done to ensure that larger positions are preferred over smaller ones; this reflects the greater prevalence of larger multiplicities seen in nature.

\begin{figure}[htbp]
    \centering
    \includegraphics[width=0.45\textwidth]{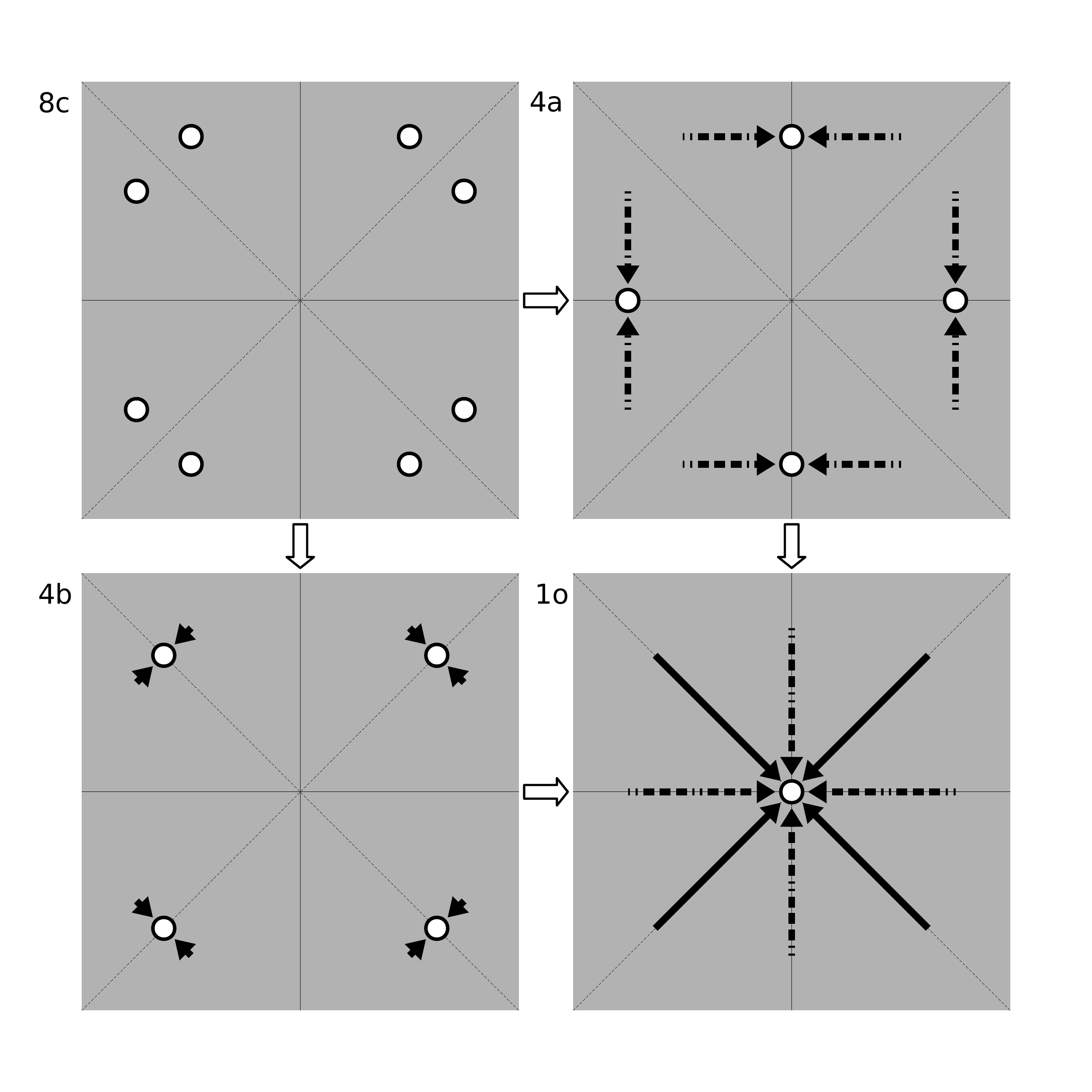}
     \caption[Wyckoff Position Merging Example]{Wyckoff Position Merging Example. Shown are possible mergings of the general position 8c of the 2D point group 4mm. Moving from 8c to 4b (along the solid arrows) requires a smaller translation than from 8c to 4a (along the dashed arrows). Thus, if the atoms in 8c were too close together, PyXtal would merge them into 4b instead of 4a. The atoms could be further merged into position 1o by following the arrows shown in the bottom right image.}
    \label{fig:WyckoffMerging}
\end{figure}

Once a WP is chosen, a random 3-vector between (0, 0, 0) and (1, 1, 1) is created. We call this the generating point for the WP. Using the closest projection of this vector onto the WP (the WP being a periodic set of points, lines, or planes), one obtains a set of coordinates in real space (the atomic positions for that WP). Then, the distances between these coordinates are checked. If the atom-atom distances are all greater than a pre-defined limit, the WP is kept and the algorithm continues. If any of the distances are too small, it is an indication that the WP would not occur with the chosen generating point. In this case, the coordinates are merged together into a smaller WP, if possible. This merging continues until the atoms are no longer too close together (see Figure \ref{fig:WyckoffMerging}).

To merge into a smaller position, the original generating point is projected into each of the remaining WPs. The WP with the smallest translation between the original point and the transformed point is chosen, provided that (1) the new WP is a subset of the original one, and (2) the new points are not too close to each other. If the atoms are still too close together after all possible mergings, the WP is discarded and another attempt is made.

Once a WP is successfully filled, the inter-atomic distances between the current WP and the already-added WPs are checked. If all distances are acceptable, the algorithm continues. More WPs are then added as needed until the desired number of atoms is reached. At this point, either a satisfactory structure has been generated, or the generation has failed. If the generation fails, then choosing either smaller distances tolerances or a larger volume factor might increase the chances of success. However, altering these quantities too drastically may result in less realistic crystals. Common sense and system-specific considerations should be applied when adjusting these parameters.

\subsubsection{Distance Checking}

To produce structures with realistic bonds and bond lengths, the generated atoms should not be too close together. In PyXtal, this means that, by default, two atoms should be no closer than the covalent bond length between them. However, for a given application, the user may decide that shorter or longer cutoff distances are appropriate. For this reason, PyXtal has a custom \textit{tolerance matrix} class which allows the user to define the distances allowed between any two atomic species. There are also options to use the metallic bond lengths, or to simply scale the allowed distances by some factor.

Because crystals have periodic symmetry, any point in a crystal actually corresponds to an infinite lattice of points. Likewise, any separation vector between two points actually corresponds to an infinite number of separation vectors. For the purposes of distance checking, only the shortest of these vectors are relevant. When a lattice is non-Euclidean, the problem of finding shortest distances with periodic boundary conditions is non-trivial, and the general solution can be computationally expensive \cite{LatticeProblem}. So instead, PyXtal uses an approximate solution based on assumptions about the lattice geometry:

For any two given points, PyXtal first considers only the separation vector which lies within the ``central'' unit cell spanning between (0, 0, 0) and (1, 1, 1). For example, if the original two (fractional) points are (-8.1, 5.2, -4.8) and (2.7, -7.4, 9.3), one can directly obtain the separation vector (-10.8, 12.6, -14.1). This vector lies outside of the central unit cell, so we translate by the integer-valued vector (11.0, -12.0, 15.0) to obtain (0.2, 0.6, 0.9), which lies within the central unit cell. PyXtal also considers those vectors lying within a 3$\times$3$\times$3 supercell centered on the first vector. In this example, these would include (1.2, 1.6, 1.9), (-0.8, -0.4, -0.1), (-0.8, 1.6, 0.9), etc. This gives a total of 27 separation vectors to consider. After converting to absolute coordinates (by dotting the fractional vectors with the cell matrix), one can calculate the Euclidean length of each of these vectors and thus find the shortest distance.

\begin{figure}[htbp]
    \centering
    \includegraphics[width=0.4\textwidth]{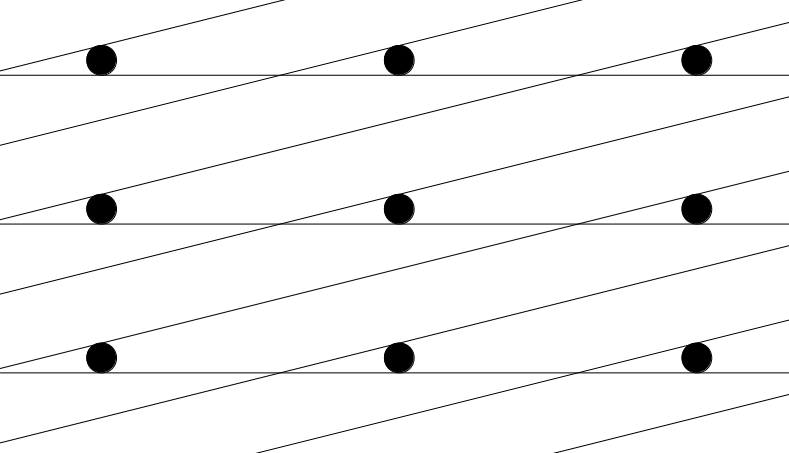}
    \caption[Distorted Unit Cell]{Distorted Unit Cell. Due to the cell's high level of distortion, the closest neighbors for a single point lie more than two unit cells away. In this case, the closest point to the central point is located two cells to the left and one cell diagonal-up. To find this point using PyXtal's distance checking method, a 5$\times$5$\times$5 unit cell will be created. For this reason, a limit is placed on the distortion of randomly generated lattices.}
    \label{fig:SkewedUnitCell}
\end{figure}

Note that this does not work for certain vectors within some highly distorted lattices (see Figure \ref{fig:SkewedUnitCell}). Often, the shortest Euclidean distance is accompanied by the shortest fractional distance, but whether this is the case or not depends on how distorted the lattice is. However, because randomly generated lattices in PyXtal are required to have no angles smaller than 30 degrees or larger than 150 degrees, this is not an issue.

For two given sets of atoms (for example, when cross-checking two WPs in the same crystal), one can calculate the shortest inter-atomic distances by applying the above procedure for each unique pair of atoms. This only works if it has already been established that both sets on their own satisfy the needed distance requirements.

Thanks to symmetry, one needs not calculate every atomic pair between two WPs. For two WPs, A and B, it is only necessary to calculate either (1) the separations between one atom in A and all atoms in B, or (2) one atom in B and all atoms in A. This is because the symmetry operations which duplicate a point in a WP also duplicate the separation vectors associated with that point. This is also true for a single WP; for example, in a Wyckoff position with 16 points, only 15 (the number of pairs involving one atom) distance calculations are needed, as opposed to 120 (the total number of pairs). This can significantly speed up the calculation for larger WPs.

For a single WP, it is necessary to calculate the distances for each unique atom-atom pair after symmetry reduction, but also for the lattice vectors for each atom by itself. Since the lattice is the same for all atoms in the crystal, this check only needs to be performed on a single atom of each specie. For atomic crystals, this just means ensuring that the generated lattice vectors are sufficiently long.

\begin{figure}[htbp]
    \centering
    \includegraphics[width=0.45\textwidth]{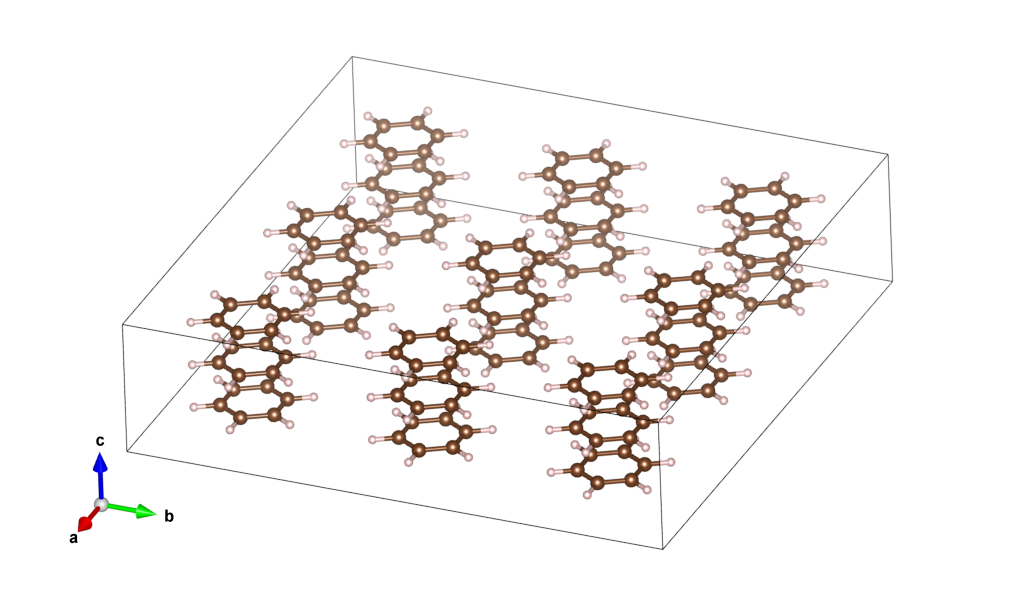}
    \caption[Dependence of shortest distances on molecular Orientation]{Dependence of shortest distances on molecular orientation. Rotation of the molecules about the $a$ or $b$ (but not the $c$) axes would cause the benzene molecules to overlap. PyXtal checks for overlap whenever a molecular orientation is altered.}
    \label{fig:BenzeneBox}
\end{figure}

For molecules, the process is slightly more complicated. Depending on the molecule's orientation within the lattice, the inter-atomic distances can change. Additionally, one must calculate the distances not just between molecular centers, but between every unique atom-atom pair. This increases the number of needed calculations in rough proportion to the square of size of the molecules. As a result, this is typically the largest time cost for generation of molecular crystals. The issue of checking the lattice is also dependent on molecular orientation. Thus, the lattice must be checked for every molecular orientation in the crystal. To do this, the atoms in the original molecule are checked against the atoms in periodically translated copies of the molecule. Here, standard atom-atom distance checking is used. While several approximate methods for inter-molecular distance checking exist, their performance is highly dependent on the molecular shape and number of atoms. The simplest method is to model the molecule as a sphere, in which case only the center-center distances are needed. This works well for certain molecules, like buckminsterfullerene, which have a large number of atoms and are approximately spherical in shape. But a spherical model works poorly for irregularly shaped molecules like benzene (see Figure \ref{fig:BenzeneBox}), which may have short separations along the perpendicular axis, but must be further apart along the planar axes. We provide spherical distance checking as an option for the user, but direct atom-atom distance checking is used by default.

\subsubsection{Molecular Orientations}

In crystallography, atoms are typically assumed to be spherically symmetric point particles with no well-defined orientation. Since the object occupying a crystallographic WP is usually an atom, it is further assumed that the object's symmetry group contains the WP's site symmetry as a subgroup. If this is the case, the only remaining condition for occupation of a WP is the location within the unit cell. However, if the object is instead a molecule, then the WP compatibility is also determined by orientation and shape.

To handle the general case, one must ensure that the object is (1) sufficiently symmetric, and is (2) oriented such that its symmetry operations are aligned with the Wyckoff site symmetry. The result is that objects with different point group symmetries are only compatible with certain WPs. For a given molecule and WP, one can find all valid orientations as follows:

1. Determine the molecule's point group and point group operations. This is currently handled by Pymatgen's built-in \textit{PointGroupAnalyzer class} \cite{pymatgen}, which produces a list of symmetry operations for the molecule.

2. Associate an axis to every symmetry operation. For a rotation or improper rotation, we use the rotational axis. For a mirror plane, we use an axis perpendicular to the plane. Note that inversional symmetry does not add any constraints, since the inversion center is always located at the molecule's center of mass.

3. Choose up to two non-collinear axes from the site symmetry and calculate the angle between them. Find all conjugate operation pairs (with the same order and type) in the molecular point symmetry with the same angle between the axes, and store the rotation which maps the pairs of axes onto each other. For example, if the site symmetry were mmm, then we could choose two reflectional axes, say the x- and y- axes or the y- and z- axes. Then, we would look for two reflection operations in the molecular symmetry group. If the angle between these two operation axes is also 90 degrees, we would store the rotation that maps the two molecular axes onto the Wyckoff axes. We would also do this for every other pair of reflections with 90 degrees separating them.

4. For a given pair of axes, there are two rotations which can map one onto the other. There is one rotation which maps the first axis directly onto the second and another rotation which maps the first axis onto the opposite of the second axis. Depending on the molecular symmetry, the two resulting orientations may or may not be symmetrically equivalent. So, using the list of rotations calculated in step 3, remove redundant orientations which are equivalent to each other.

5. For each found orientation, check that the rotated molecule is symmetric under the Wyckoff site symmetry. To do this, simply check the site symmetry operations one at a time by applying each operation to the molecule and checking for equivalence with the untransformed molecule.

6. For the remaining valid orientations, store the rotation matrix and the number of degrees of freedom. If two axes were used to constrain the molecule, then there are no degrees of freedom. If one axis is used, then there is one rotational degree of freedom, and we store the axis about which the molecule may rotate. If no axes are used (because there are only point operations in the site symmetry), then there are three (stored internally as two) degrees of freedom, meaning the molecule can be rotated freely in 3 dimensions.

PyXtal performs these steps for every WP in the symmetry group and stores the nested list of valid orientations. When a molecule must be inserted into a WP, an allowed orientation is randomly chosen from this list. This forces the overall symmetry group to be preserved since symmetry-breaking WPs do not have any valid orientations to choose from. The above algorithm is particularly useful to generate molecular crystals with non-integer number of molecules in the asymmetric unit, which occur frequently for molecules with high point group symmetry.

One important consideration is whether a symmetry group will produce inverted copies of the constituent molecules. In many cases, a chiral molecule's mirror image will possess different chemical or biological properties \cite{chirality}. For pharmaceutical applications in particular, one may not want to consider crystals containing mirror molecules. By default, PyXtal does not generate crystals with mirror copies of chiral molecules. The user can choose to allow inversion if desired.

\subsection{Structure Modulation}
In many applications, it is often necessary to derive a new structure from the parent structure models. The main concept is to start from a simple, highly symmetrical crystal structure and to derive more complicated structures by applying a fairly small distortion or chemical substitution. Noticeable examples include non-reconstructive phase transitions and catalogue of structural prototype. In structure prediction methods such as simulated annealing and evolutionary algorithms, the generation of new structures are usually obtained by applying a random perturbation on both lattice vectors and atomic coordinates. In this context, the introduction of symmetry relation can reduce the search space greatly. PyXtal provides two ways of structure manipulation based on the symmetry constraints.

\subsubsection{Perturbation on asymmetric unit}
The most straightforward way to preserve the crystal symmetry is to perturb only the atoms in the asymmetric unit. For atomic crystals, we only apply a random perturbation to each Wyckoff site if it has free coordinates. For molecular crystals, the whole molecule can be reoriented along the allowed rotational axes. If the molecule has flexible rotors, additional perturbation on these dihedral angles can be applied as well. Optionally, the cell parameters can be changed as well based on the symmetry constraints. Applying the symmetry-preserved mutation can be used to effectively search for molecular crystals in a constrained space group symmetry.

\subsubsection{Group-subgroup transition}
\begin{figure}[htbp]
    \centering
    \includegraphics[width=0.48\textwidth]{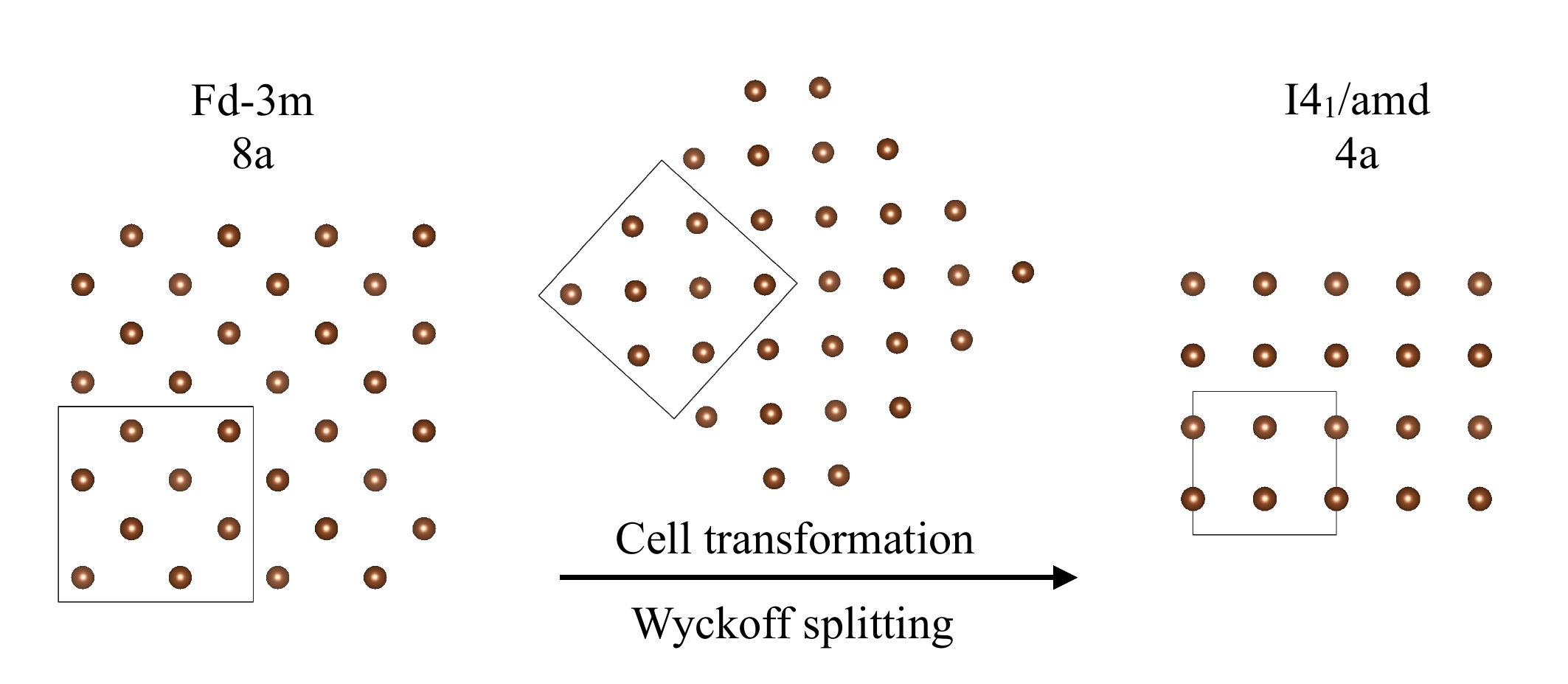}
    \caption{A schematic example to illustrate group-subgroup transition from $Fd$-$3m$ (8a) to $I4_1/amd$ (4a).}
    \label{fig:subgroup}
\end{figure}

When one crystal is converted to another by a phase transition, the symmetries of the crystal structures are usually related. The so called group-subgroup has been well discussed mathematically. It is possible to list all possible subgroup types for every space-group type and to specify the subgroups in a general way by formulae. The international crystallography volume \cite{wondratschek2006symmetry}, as well as the online Bilbao Crystallographic Server \cite{aroyo2006bilbao}, have provided the symmetry relations between a given space group $G$ and its possible maximal translationengleiche ($t$) and klassengleiche ($k$) subgroups $H$. Ideally, to complete the transition from $G$ to $H$, one needs to know the cell transformation matrix, as well as the Wyckoff splitting scheme. For a given crystal structure, PyXtal allows to either systematically extract all possible transformation (subject to a cutoff index of symmetry reduction), or randomly pick one possible transformation path between $G$ and $H$. Figure \ref{fig:subgroup} shows an example to illustrate the transition from a structure occupying 8a in $Fd$-$3m$ to another structure occupying 4a in $I4_1/amd$ symmetry.

\section{Dependencies}
\label{dependence}
All of the code is written in Python 3. Like many other Python packages, it relies on several external libraries. Numpy \cite{numpy}, Scipy \cite{scipy} and Pandas \cite{pandas} are required for the general purposes of scientific computing and data processing. In addition, two materials science libraries, Pymatgen \cite{pymatgen} and Spglib \cite{spglib}, were used to facilitate the symmetry analysis. Optionally, the code provides an interface with Openbabel \cite{openbabel} if the users wants to import the molecules from additional file formats other than the plain xyz format. An ASE \cite{ASE} interface is also enabled if the user wants to do further structure analysis such as structure manipulation or geometry optimization based on ASE.

\section{Example Usages}
\label{usage}
PyXtal can be either used as a binary executable or stand-alone library for use in Python scripts. Below we introduce the basic usages in brief.  
\label{example}
\subsection{Command line utilities}
Currently, several utilities are available to access the different functionality of PyXtal. They include:
\begin{enumerate}
    \item pyxtal\_symmetry.py
    \item pyxtal\_main.py
    \item pyxtal\_test.py
\end{enumerate}

First, the users are advised to run the pyxtal\_test.py to quickly test if all modules are working correctly after the installation. The rest of the utilities are designed for different analysis purposes. 

The pyxtal\_symmetry.py utility allows one to easily access the symmetry information for a given symmetry group using either the group name or international number.
\begin{lstlisting}[language=Python, caption=Example usage of the pyxtal\_symmetry.py utility.]
$ pyxtal_symmetry.py -s 64
-- Spacegroup --# 64 (Cmce)--
16g	site symm: 1
8f	site symm: m..
8e	site symm: .2.
8d	site symm: 2..
8c	site symm: -1
4b	site symm: 2/m..
4a	site symm: 2/m..
\end{lstlisting}

The pyxtal\_main.py can be used to directly generate one trial structure based on the given symmetry group and chemical composition. Below, we give the example scripts to generate different types of symmetric objects, including 
\begin{enumerate}
    \item a random C60 cluster with $I_h$ point group symmetry; 
    \item a trial diamond structure with \textit{Fd-3m} space group symmetry;
    \item a crystal of two C60 molecules per primitive unit cell with $Cmc2_1$ symmetry
\end{enumerate}

\begin{lstlisting}[language=Python, caption=Example usages of the pyxtal\_main utility.]
$ pyxtal_main.py -e C -n 60 -d 0 -s Ih
$ pyxtal_main.py -e C -n 2 -s 227
$ pyxtal_main.py -m -e C60 -n 4 -s 36
\end{lstlisting}

The generated structures will be saved to text files in cif format for crystals and xyz format for clusters.

\subsection{Structure modulation and analysis}
\begin{figure}[htbp]
    \centering
    \includegraphics[width=0.48\textwidth]{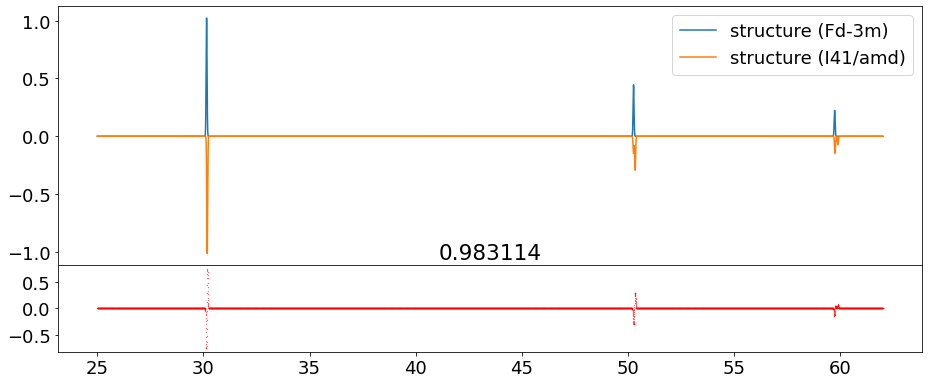}
    \caption{The comparison of simulated X-ray diffraction patterns between the original structure and the perturbed structure from the group-subgroup relation.}
    \label{fig:xrd}
\end{figure}

In addition to structure generation and manipulation, PyXtal also provides several other utilities, such as XRD analysis. Below, we give the example scripts to (1) generate a cubic crystal, (2) perturb the structure to lower the crystal symmetry by following the group-subgroup relation, and (3) compare two PXRD between two structures. Finally, Figure \ref{fig:xrd} displays the simulated XRDs between two structures, where the similarity between two XRDs are also given according to the correlation function as suggested previously \cite{de2001generalized, habermehl2014structure}. Note the split of XRD peaks.

\begin{lstlisting}[language=Python, caption={Example usages of PyXtal's structural manipulation and analysis utilities}, label=l2]
from pyxtal import pyxtal
from pyxtal.XRD import Similarity

# generate a random crystal
C1 = pyxtal()
C1.from_random(3, 227, ['C'], [8])

# perturbation without changing the symmetry
# C2 = C1.copy()
# C2.apply_perturbation()

# lower the symmetry from cubic to tetragonal
C2 = C1.subgroup(H=141, once=True)

# Compute the XRD
xrd1 = C1.get_XRD()
xrd2 = C2.get_XRD()

# Compare two structures by XRD
p1 = xrd1.get_profile()
p2 = xrd2.get_profile()
s = Similarity(p1, p2, x_range=[15, 90])
s.show(filename='xrd-comparison.png')
\end{lstlisting}

\subsection{Structure Prediction}
PyXtal allows the user to generate random crystal structures with given symmetry constraints. There are several parameters which can be specified, but only a few are necessary. Below is an example script to generate 100 random clusters for 36 carbon atoms.

\begin{lstlisting}[language=Python, caption=A Python script to generate 100 random C36 clusters]
from pyxtal import pyxtal
from random import choice

pgs = range(1, 33)
clusters = []
for i in range(100):
    while True:
        pg = choice(pgs)
        cluster = pyxtal()
        cluster.from_random(0, pg, ['C'], [36], force_pass=True)
        if cluster.valid:
            clusters.append(cluster)
            break
\end{lstlisting}

With the generated structures, one can perform further analysis such as geometry optimization and powder X-ray diffraction pattern simulation. PyXtal also provides the preliminary modules for such tasks. Alternatively, the trial structures can be easily adapted to the structural objects for other libraries, such as ASE \cite{ASE} or Pymatgen \cite{pymatgen}, or be dumped to text files in cif, xyz, or POSCAR format. More examples can be found in the online documentation \url{https://pyxtal.readthedocs.io}.  

\section{Applications}
\label{examples}
Our primary purpose for developing PyXtal is to provide more likely trial structures to solve the structural determination problem. It can be useful for at least two cases. First, one can generate the trial structures based on the partial information determined from experiment (e.g., unit cell, symmetry, composition). Secondly, it can be used to determine the ground state structure in a first-principle manner based on global optimization. It has been shown \cite{Oganov-JCP-2006, CALYPSO, randspg} that by beginning with already-symmetric structures, fewer attempts are needed to find the global energy minimum. To demonstrate the general utility of pre-symmetrization, we performed a number of benchmarks for different systems. Below we give two examples for the global structural search on the low-energy Lennard-Jones (LJ) clusters and carbon/silicon allotropes.

\subsection{Clusters with empirical Lennard-Jones potential}
\begin{figure}[htbp]
    \centering
    \includegraphics[width=0.45\textwidth]{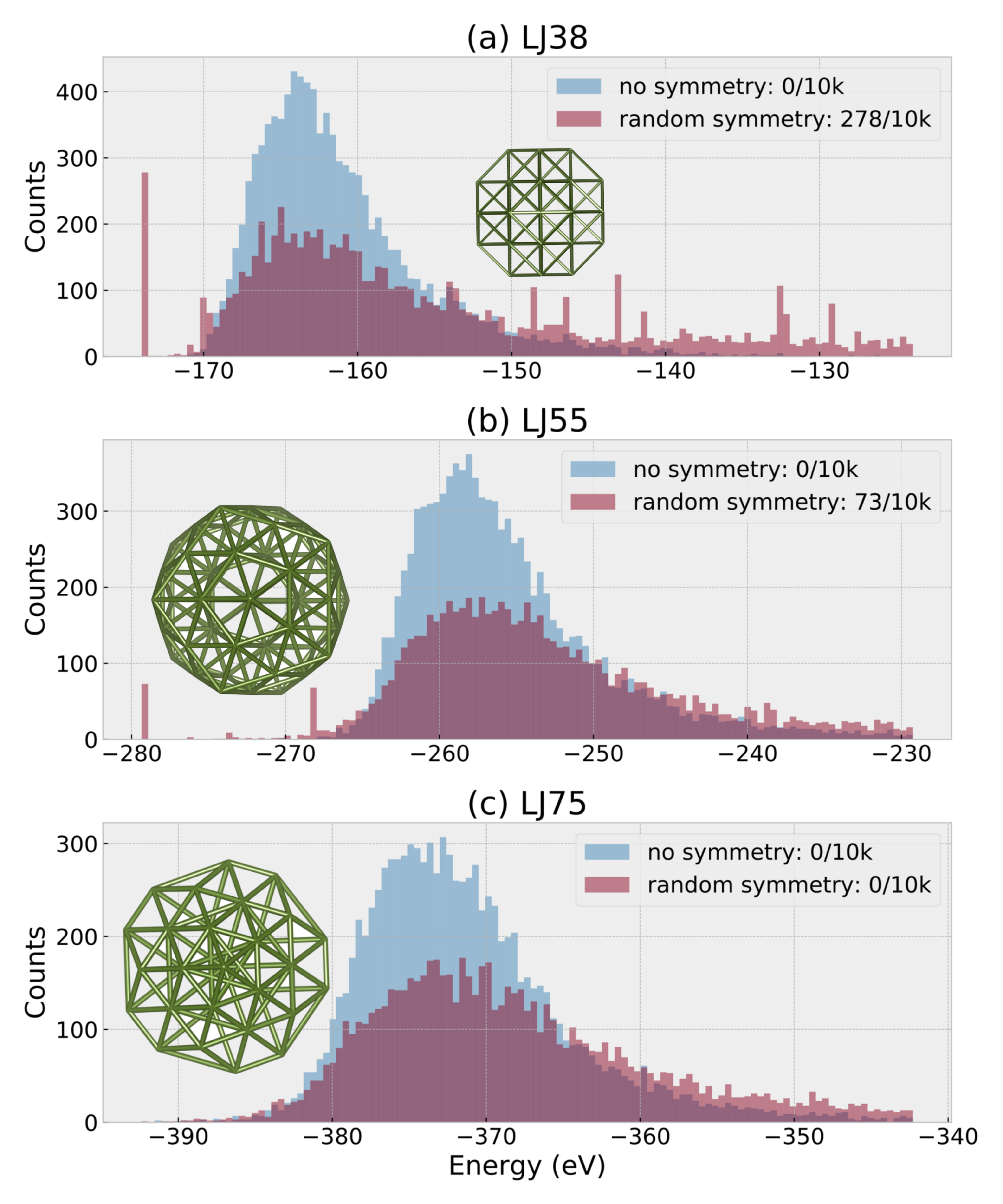}
    \caption{Energy distribution for Lennard Jones clusters with the sizes of (a) 38, (b) 55 and (c) 75. The insets are the corresponding ground state geometries.}
    \label{fig:LJ}
\end{figure}

Finding the ground state of LJ clusters of given size is an established benchmark for global optimization methods \cite{LJClusters}. Here, it shows that local optimization, combined with randomly generated symmetric clusters, is sufficient to solve the problem with small sizes of LJ clusters. For the purposes of this benchmark, we focus on three cluster sizes, namely 38, 55, and 75. For each cluster size, 20,000 structures were generated: 10,000 with no pre-defined symmetry and 10,000 with symmetry chosen randomly from among PyXtal's 56 built-in point groups. A potential of $ 4(\frac{1}{r^{12}} - \frac{1}{r^6}) $ was assigned to each atom-atom pair. Each structure was locally optimized using the conjugate gradient (CG) method in SciPy's \textit{optimize.minimize} function \cite{scipy}. As shown in Figure \ref{fig:LJ}, the ground state was found much more frequently when the initial structures possessed some point group symmetry. With pre-symmetrization, the ground state was found 278 times for size 38 clusters, 73 times for size 55, and 1 time for size 75. Without pre-symmetrization, the ground state was not found at all. Though the numbers of hits on the ground states may change in another run, the statistical rule still holds. Second, while the ground state is found more frequently with pre-symmetrization, the average energy is higher. This is because pre-symmetrization spans the possible structure space more effectively, while purely random structures are more clustered around a specific energy range. 

\begin{figure}[htbp]
    \centering
    \includegraphics[width=0.45\textwidth]{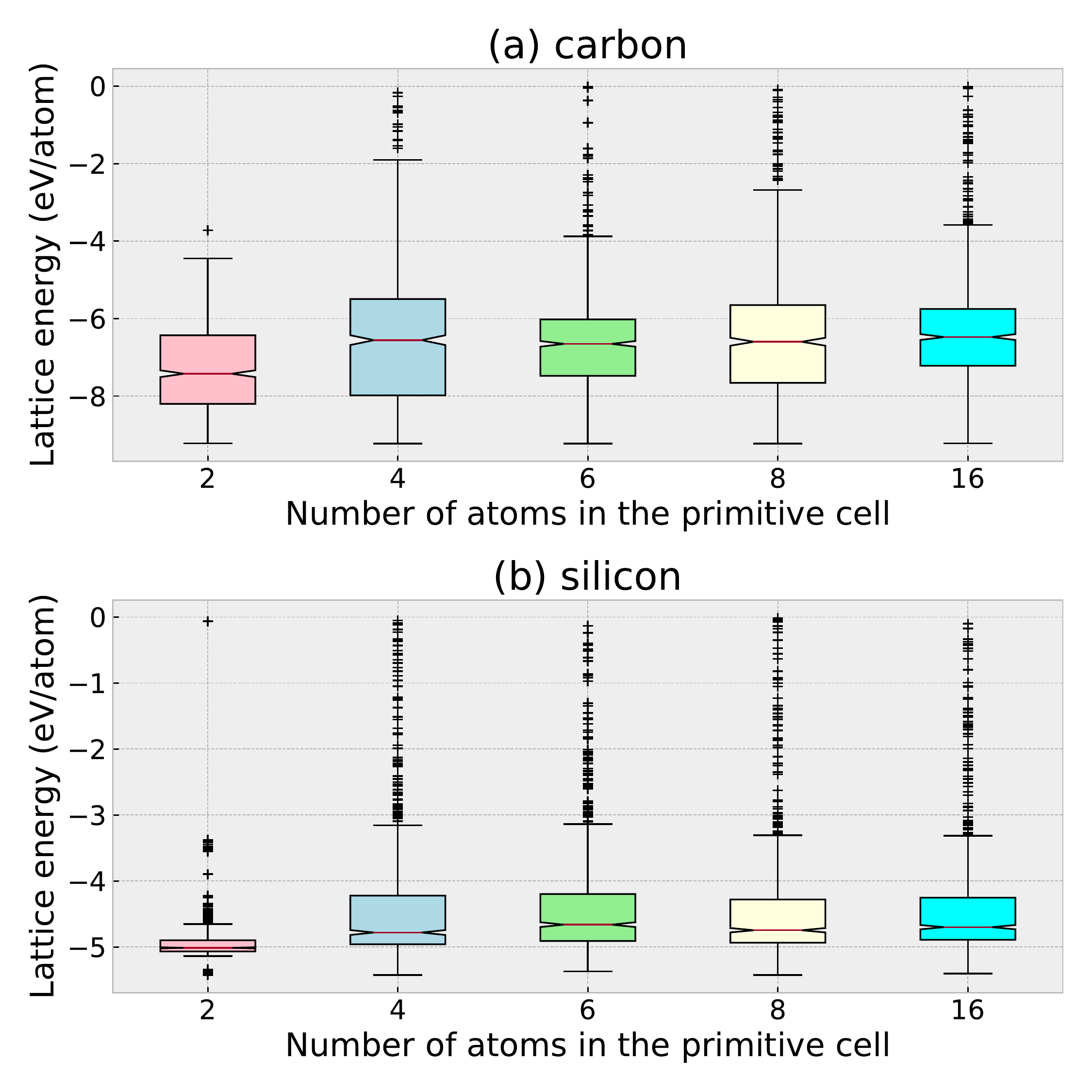}
    \caption{The box and whisker plots for the energy distribution of the randomly generated (a) carbon and (b) silicon crystals with 2, 4, 6, 8, 16 atoms per primitive unit cell.}
    \label{fig:boxplot}
\end{figure}

\subsection{Carbon and silicon crystals with ab-initio calculations}
We also combined PyXtal with ab-initio codes to search for the elemental allotropes of carbon and silicon at 0 K and ambient pressure. 1000 random structures each were generated for 2, 4, 6, 8, and 16 atoms in the primitive unit cell. A random space group between 2 and 230 was chosen for each structure. This gave a total of 5000 structures for each element. Each structure was optimized using the PBE-GGA functional \cite{PBE-PRL-1996} as implemented in the VASP code \cite{VASP3, VASP4}, following a multiple-step strategy from low, normal, to accurate precision. The final geometries were then calculated with an energy cutoff of 600 eV and 0.15 K-spacing. For carbon, the expected structures of diamond and graphite were found frequently in each run, as well as londsdaelite, $sp3$ carbon with various ring topologies, and various multi-layer graphite-like structures. Similarly, our simulation on silicon yielded the ground state of cubic diamond structures for each of the runs with different numbers of atoms per primitive unit cell, demonstrating that adding symmetry constraints is beneficial to quickly identify the low-energy structures with high symmetry. Moreover, it is again interesting to analyze the energy distribution of the randomly generated structures as shown in Figure \ref{fig:boxplot}. For both carbon and silicon, the energy landscape appears to be narrower for size-2 primitive cells. It appears that beyond about 4, the number of atoms in the primitive cell has little influence on the energy distribution. This again suggests that pre-symmetrization is an effective means to prevent the clustering of glassy structures found in pure random generations for large systems \cite{Lyakhov-CPC-2013}. Therefore, pre-symmetrization provides a better choice for global energy optimization. In addition, pre-symmetrization can provide a more diverse dataset for training machine learning force fields \cite{Deringer-PRL-2018, Boron-PRB-2019}.

\section{Conclusion}
\label{conclusion}

In this manuscript, we present a software package PyXtal. The core features of PyXtal have been highlighted, with further documentation available online\footnote{\url{https://pyxtal.readthedocs.io}}. In PyXtal, the symmetry constraints are further refined in three ways. The first is a merging algorithm \cite{zhu2012} which controls the distribution of WPs through statistical means. The second is a new algorithm for placing molecules into special WPs. Last is the structure modulation by the observation of the symmetry relation. This allows for more realistic and complex structures to be generated by keeping the symmetry as high as possible. PyXtal is not a complete structure prediction package; it only generates the trial structures with a given symmetry group. Other tools exist that perform structure generation and other steps in the CSP process \cite{Lyakhov-CPC-2013, Pickard-JPCM-2011, XtalOpt, CALYPSO}. The main goals in developing PyXtal are as follows: 1) to develop a free, open-source Python package for the materials science community, 2) to handle the generation of symmetric structures described by different symmetry groups from 0D to 3D, 3) to handle molecular WPs in a generalized manner, 4) to provide a tool to analyze the symmetry relation. We also demonstrated that using the pre-symmetrized structures as the starting seeding structures can effectively improve the success rate of finding the low energy configuration. As such, PyXtal can be interfaced with other structure prediction codes that require the generation of trial structures. Access to the source code and development information are available on the GitHub page at \url{https://github.com/qzhu2017/PyXtal}. The code is currently under version 0.1.5 at the time of writing. It is expected to update frequently. Further development and application of the mathematical background should enable more complex structure types to be studied in the future.

\section*{Acknowledgments}
We acknowledge the NSF (I-DIRSE-IL: 1940272) and NASA (80NSSC19M0152) for their financial supports. The computing resources are provided by XSEDE (TG-DMR180040).

\bibliographystyle{elsarticle-num}
\bibliography{ref.bib}
\end{document}